\documentclass[
    ,final            % use final for the camera ready runs
  ]
  {aipproc}

\layoutstyle{8x11double}

\begin{document}

\title{Furnishing the Galaxy with Pulsars}


\author{Paul Kiel}{
  address={Center for Astrophysics and Supercomputing, Swinburne University of Technology, Hawthorn, Victoria, 3122, Australia}
}

\author{Jarrod Hurley}{
%  address={<common address for author2 and author3>}
}

\author{Matthew Bailes}{
%  address={<common address for author2 and author3>}
%  ,altaddress={<author1 address>} % additional visiting address
}
\author{James Murray}{
%  address={<common address for author2 and author3>}
}

\begin{abstract}
The majority of pulsar population synthesis studies performed to 
date have focused on isolated pulsar evolution. 
Those that have incorporated pulsar evolution within binary systems 
have tended to either treat binary evolution poorly of evolve the 
pulsar population in an ad-hoc manner. 
Here we present the first model of the Galactic field pulsar population 
that includes a comprehensive treatment of both binary and pulsar evolution.
Synthetic observational surveys mimicking a variety of radio telescopes 
are then performed on this population. 
As such, a complete and direct comparison of model data with observations 
of the pulsar population within the Galactic disk is now possible.
The tool used for completing this work is a code comprised of three 
components: stellar/binary evolution, Galactic kinematics and survey 
selection effects. Here we give a brief overview of the method and 
assumptions involved with each component. 
Some preliminary results are also presented as well as plans for 
future applications of the code.
\end{abstract}

%%%%%%%%%%%%%%%%%%%%%%%%%%%%%%%%%%%%%%%%%%%%%%%%%%%%%%%%%%%%%%%%%%%
%%
%% The below \maketitle command inserts the actual front matter data.
%% It has to follow the above declarations.
%%
%%%%%%%%%%%%%%%%%%%%%%%%%%%

\maketitle

%%%%%%%%%%%%%%%%%%%%%%%%%%%%%%%%%%%%%%%%%%%%
%% MAINMATTER
%%
%%%%%%%%%%%%%%%%%%%%%%%%%%%%%%%%%%%%%%%%%%%%%%%%%%%%%%%%%%%%%%%%%%%%%%%%%%%%
%% Headings:
%%
%% The aipproc supports three heading levels, i.e., \section,
%%	\subsection, and \subsubsection.
%%%%%%%%%%%%%%%%%%%%%%%%%%%%%%%%%%%%%%%%%%%%%%%%%%%%%%%%%%%%%%%%%%%%%%%%%%%%
%% Cross-references:
%%
%% Page numbers (\pageref) and headings can NOT be referenced in the class,
%% since before being produced, no page numbers are determined.
%%
%% Tables, figures, and equeations can be referenced by using the LaTex
%% 	commands \label and \ref. For references to equation numbers, \eqref
%%	can be used, which will print "(1)" (while \ref will result in "1").
%%
%%%%%%%%%%%%%%%%%%%%%%%%%%%%%%%%%%%%%%%%%%%%%%%%%%%%%%%%%%%%%%%%%%%%%%%%%%%%
%% Lists: 
%%
%% Standard "itemize", "enumerate", etc. list environments are supported.
%%%%%%%%%%%%%%%%%%%%%%%%%%%%%%%%%%%%%%%%%%%%%%%%%%%%%%%%%%%%%%%%%%%%%%%%%%%%
%% Urls:
%%
%% \url{} command is provided for documenting URLs.
%%%%%%%%%%%%%%%%%%%%%%%%%%%%%%%%%%%%%%%%%%%%

\section{Population synthesis}
We evolve n systems by first selecting the initial parameters of each 
single star (mass) or binary (masses, period) from appropriate 
distributions, e.g.- stellar masses drawn from the Kroupa, Tout \& 
Gilmore (1993) initial mass function. 
Limits must also be chosen where we typically take $5-80 M_\odot$
and $0.1-40 M_\odot$ for the binary primary and secondary masses. 
The algorithm loops over the n systems in \textbf{binpop} which follows the 
stellar/binary evolution. 
Any systems of interest that are produced (in this case pulsars) are 
passed to \textbf{binkin} which follows the kinematic evolution 
of each system. 
The resulting population is then `surveyed' in \textbf{binsfx} which produces 
synthetic data for analysis.

\section{\textbf{binpop}: Binary and Pulsar evolution}
This component evolves single and binary stars from the zero age main 
sequence through to some given time, $t_{max}$. 
It uses the binary evolution code of Hurley, Tout \& Pols (2002) 
with the addition of pulsar evolution. 
Some features included in the binary evolutionary algorithm are tides, 
mass transfer (examples are stellar winds, roche-lobe overflow and 
common envelope), magnetic braking, orbital gravitational radiation, 
mergers and supernova velocity kicks. 
Features particular to neutron star (NS) include, standard pulsar 
magnetic field decay, accretion induced field decay and spin-up, 
magnetic braking, propeller effects and death lines.
In this body of work we focus on a particular model which makes 
the following assumptions and produced the pulsars shown in Figure 1. 
The model evolves n~$=10^7$ binary systems, where each system is evolved 
to a $t_{max}$ drawn at random from between the limits of 0 and 10 Gyr. 
If a NS forms it's initial spin period $P_0$ and magnetic field $B_0$ are 
given by $P_0 = 0.02+0.156(V_{kick}/V_{\sigma})$ [s] and 
$B_0 = 5x10^{11}+4x10^{12}\times(V_{kick}/V_{\sigma})$ [G], where $V_{kick}$ 
is randomly drawn from a Maxwellian with standard deviation 
$V_{\sigma} = 190$ km/s. 
In terms of NS magnetic field decay we assumed the standard decay time 
constant , $\tau_B = 1000$ Myr while the accretion induced decay scaling 
parameter is, $k = 7000$. 
We then ignore pulsars whose magnetic field decays to the limiting value 
of $5\times10^7$ G. 
If a NS is accreting we assume it is not visible as a pulsar and it is 
ignored, this includes those NS who were accreting material via Roche-lobe 
overflow within the previous 1000 years. 
Finally, this model also includes the Harding, Muslimov \& Zhang (2002) 
curvature radiation death line, beyond which pulsars are assumed not to 
emit beamed emission of sufficient strength to be observed.

\begin{figure}
  \includegraphics[height=.3\textheight]{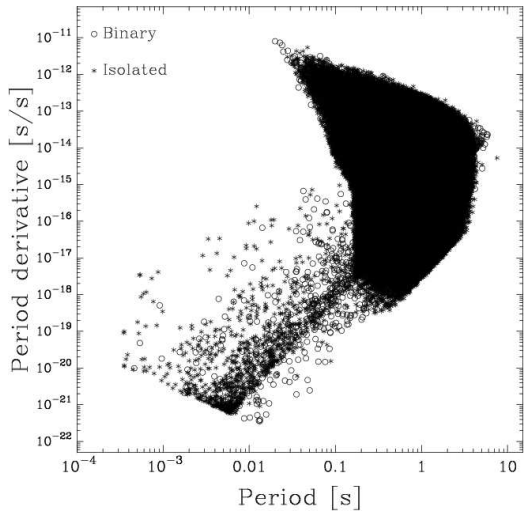}
  \caption{Spin period against spin period derivative for each pulsar
  produced in the model described to the left. Each point is one pulsar,
  circles represent binary pulsars while stars are isolated pulsars.}
\end{figure}

\section{\textbf{binkin}: Galactic kinematics}
In this component single stars and binary stars are evolved dynamically 
within the Galactic gravitational potential, which we model as given by 
Kuijken \& Gilmore (1989). 
This is our first and simplest model of the Galaxy and includes the disk, 
nucleus and bulge. 
\textbf{Binkin} also takes into account supernovae, which may impart some 
momentum onto a single star or binary system or, in fact, disrupt a 
binary system. 
Every system has it's orientation randomised with respect to the 
galactic plane. 
The code has been tested and is reversible. We note that this is 
not an n-body code, each system is evolved independently. 
Each of the pulsars in Figure 2 \& 3 
are evolved in \textbf{binkin} which moves 
them in the Galaxy to produce an 3-dimensional spatial distribution. % of 
%which the x-y projection is shown in Figure 3. 
In this model the stars are assumed to be born in a disk with a 
radius of 15 kpc and a height above the plane of 50pc.

%\begin{figure}
%  \includegraphics[height=.3\textheight]{SimGalXYmontrealPoster}
%  \caption{Simulation of the Galactic ulsar distribution.
%  Each point represents one pulsar.}
%\end{figure}

\section{\textbf{binsfx}: Synthetic surveys}
We now consider the final component of our work, \textbf{binsfx}, which uses 
both the pulsar data from \textbf{binpop} and the Galactic positional data 
from \textbf{binkin} to produce synthetic surveys. 
These synthetic  surveys are based on the Parkes Multibeam (PMB), 
Swinburne intermediate latitude (SIL) and Swinburne extended (SE) 
surveys along with the Burgay et al. (Betal) survey. 
We also make use of the GMRT, which has not yet been used for an 
extended and large pulsar survey project.
In modeling the surveys it is important that we also take into 
account the associated pulsar selection effects. 
Within the model presented here (Figures 2 \& 3) we have considered beaming 
of the pulses away from our line of sight, scattering of the image 
(multi-path propagation), dispersion (a lag between different wavelengths) 
and sampling of the pulse with a finite time constant. 
We also need to assume some form of the pulsar luminosity and we 
do this using, 
$L = 0.1 \times 10^{0.1\chi} \times P^{\alpha}\times \frac{dp}{dt}^{\beta}$ 
where $\alpha = -2.0+0.1\chi$, $\beta = 0.0+0.1\chi$ and $\chi$ is a 
Gaussian random number. 
We assume a spectral index of the form $-0.8\chi$.
Once we check whether each pulsar produced above (depicted by Figure 1) 
is within the correct region of the Galaxy for the particular survey 
and is close enough that our assumed telescope sensitivity is less than 
the calculated pulsar luminosity (at the solar system) then we may start 
to produce plots like Figures 2 \& 3. 
Note that Figure 2 is a reproduction of Figure 1 after passing the 
population through \textbf{binkin} and \textbf{binsfx}. 
We may now compare these to the observations. 
For our simulated surveys of PMB, SIL, SE and Betal we find 
1180, 105, 54 and 99 pulsars respectively. 
With the above assumptions we predict that the GMRT could see 366 pulsars.

\begin{figure}
  \includegraphics[height=.3\textheight]{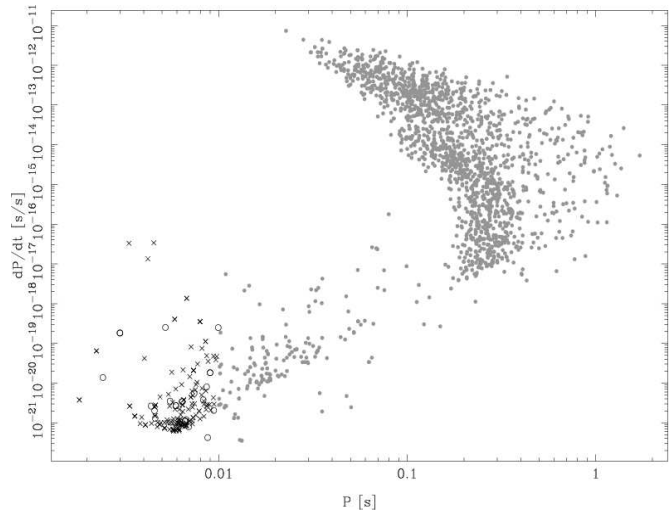}
  \caption{Synthetically observed pulsars from the Galactic pulsar population
    shown Figure 1. Each point is one pulsar, crosses are isolated millisecond 
    pulsars while circles are binary millisecond pulsars.}
\end{figure}

\begin{figure}
  \includegraphics[height=.3\textheight]{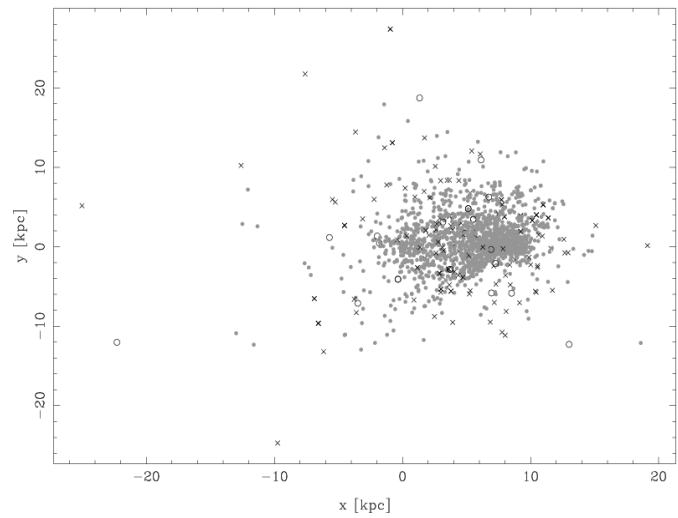}
  \caption{The Galactic x-y positions of those synthetically observed pulsars
    shown in Figure 2. Each point is one pulsar, crosses are isolated 
    millisecond pulsars while circles are binary millisecond pulsars.}
\end{figure}
\section{Future work}

The results shown here are a guide to what we can now  
achieve when modelling pulsars within the Galaxy. 
Being able to directly compare the results with observation, 
although not new, is a step forward when simulating with such 
detailed stellar and binary evolution codes.
For example, the number of pulsars the above simulated surveys 
`see' are obviously only estimates, however, through comparison 
with pulsar surveys we are now able to better constrain the 
complete underlying pulsar luminosity law. 
This work will also help in constraining binary and pulsar theory 
by allowing us to compare the spatial distributions of differing 
system types (for example see Figure 3) and the numbers and birth 
rates of these systems as directly compared to observations. 
Note that the binary evolution algorithm is not restricted to 
pulsars and any star/binary of interest can be investigated. 
This work will also help predict what next generation telescopes, 
such as the Square Kilometer Array, should see.

%%%%%%%%%%%%%%%%%%%%%%%%%%%%%%%%%%%%%%%%%%%%%%%%
%% BACKMATTER
%%%%%%%%%%%%%%%%%%%%%%%%%%%%%%%%%%%%%%%%%%%%%%%%

\begin{theacknowledgments}
PK thanks Swinburne university of Technology for travel and 
accommodation assistance.
\end{theacknowledgments}

%% The code below assumes that BibTeX is used. Compliant BibTex styles
%% are aipproc (for use with natbib) and aipprocl (if natbib is missing
%% at the site).
%%
%% Please run "bibtex \jobname" to obtain the bibliography and 
%% then re-run LaTeX twice to fix the references!
%%
%% When referring to citations in the text, in quare brackets [] show
%% the number in order of appearance. References in the References
%% section are listed in the same numerical order.
%%%%%%%%%%%%%%%%%%%%%%%%%%%%%%%%%%%%%%%%%%%%%%%%%



%\bibliography{sample}

%%%%%%%%%%%%%%%%%%%%%%%%%%%%%%%%%%%%%%%%%%%%%%%%%
%% If the bibliography is
%% produced without BibTeX, comment out the above lines, use
%% \begin{thebibliography}{widest-label} environment to hold 
%% the list of references and 
%% \bibitem{label} command to start a bibliographical entry having
%% the "label" for use in \cite commands.
%%


\end{document}